\documentclass[
twocolumn,
reggeiaps,prd,
nofootinbib,
superscriptaddress,
showpacs,ligh
tightenlines,
]{revtex4}
\usepackage{amsmath}
\usepackage{amssymb}
\usepackage{bm}
\usepackage{color,graphicx}

\begin{document}

\title{Angular momentum sum rule for spin one hadronic systems}

\author{Swadhin K. Taneja}   
\email{taneja@skipper.physics.sunysb.edu}
\affiliation{Stony Brook University - Department of Physics and Astronomy,
Stony Brook, NY 11776 - USA}

\author{Kunal Kathuria}
\email{kk7t@virginia.edu}
\affiliation{University of Virginia - Physics Department,
382, McCormick Rd., Charlottesville, Virginia 22904 - USA}

\author{Simonetta Liuti }
\email{sl4y@virginia.edu}
\affiliation{University of Virginia - Physics Department,
382, McCormick Rd., Charlottesville, Virginia 22904 - USA} 

\author{Gary R.~Goldstein} 
\email{gary.goldstein@tufts.edu}
\affiliation{Department of Physics and Astronomy, Tufts University, Medford, MA 02155 USA.}


\begin{abstract}
We derive a sum rule for the total  quark angular 
momentum of a spin-one hadronic system within a gauge invariant decomposition of the hadron's spin.
We show that the total angular momentum can be measured through deeply virtual Compton scattering experiments using 
transversely polarized deuterons. 
\end{abstract}
\maketitle
A crucial, outstanding question in QCD is  the proton spin puzzle. A number of experiments performed
since the '80s, including the most recent HERMES, Jefferson Lab and Compass measurements,  
have confirmed that only about $30 \%$ of the proton spin is accounted by quarks, and that the quark contribution
is dominated by the valence component (see review in \cite{Nowak}). Current efforts, both in theory and experiment, are therefore directed towards
determining the contributions of the  Orbital Angular Momentum (OAM) of the quarks, as well as of the spin and 
OAM of the gluons. 
Sum rules were derived that relate the Energy Momentum Tensor's (EMT) form factors to the nucleon angular momentum  \cite{Jaffe:1989jz,Ji:1996ek}. 
\footnote{Alternative procedures to obtain explicit gauge invariant operators for spin and orbital angular momentum of quarks and gluons were given in \cite{Chen,Wak}. Their discussion is beyond the scope
of this paper, see however \cite{Leader:2011za}.}
In \cite{Jaffe:1989jz}, starting from the classical/canonical form of the EMT, it is possible to 
identify the four contributions  from the quark and gluon OAM and spin components. 
Of these only the 
quark and gluon spin terms appear among the observables for hard scattering processes. 
On the other side, the result derived in \cite{Ji:1996ek}, uses the symmetric, Belinfante form of the EMT  and leads to different definitions of 
the angular momentum components, $J_q=L_q+ \Delta \Sigma$, and 
$J_g$. These can, in principle, be measured through Deeply Virtual  Compton Scattering (DVCS) (see also \cite{Bakker:2004ib}). 
However, the interpretation of these components in terms of unintegrated parton angular momentum density distributions is not straightforward.
The values of the observables  will therefore differ in the two approaches \cite{BurkardtBC}.

Motivated by the challenge of the spin puzzle on one side, and by the feasibility of DVCS type experiments, 
we decided to investigate the angular momentum sum rules for hadronic systems of different spin which are provided, in practice, by nuclear targets. 
In this contribution we present a sum rule for the total angular momentum in a spin one nucleus, the deuteron. The sum rule is 
of particular relevance because it involves only one Generalized Parton Distribution (GPD), namely
%

%
%
\begin{equation}
\label{Ji_deuteron}
J_{q}=\frac{1}{2} \int dx \, x \, H_{2}^{q}(x,0,0).  
\end{equation}
$H_{2}^{q}(x,\xi,t)$'s first moment is equal to the deuteron magnetic form factor  $G_2(t) \equiv G_M(t)$  \cite{Berger:2001zb}. 
This expression can be compared to the nucleon sum rule \cite{Ji:1996ek},
\begin{equation}
\label{Ji}
J_q = \frac{1}{2} \int d x \, x \left[ H_q(x,0,0) + E_q(x,0,0) \right],
\end{equation}    
where  the first moment of the GPD sum $H_q(x,\xi,t) + E_q(x,\xi,t)$ is the nucleon magnetic form factor, $F_1(t)+F_2(t) \equiv G_M(t)$.
Similar to the proton GPD $E$,  $H_2$ does not have a forward partonic limit.

In what follows we outline the fundamental steps of the derivation. 
We start from the expression for  angular momentum in QCD, 
\begin{equation}
     J^i = \frac{1}{2}\epsilon^{ijk} 
      \int d^3x M^{0jk} \ ,
\label{e:angmom}       
\end{equation}
where the tensor $M^{0ij}$ is the angular momentum density given in terms of the symmetric, gauge-invariant, and conserved (Belinfante) EMT as $
      M^{\alpha\mu\nu} = T^{\alpha\nu} x^\mu - T^{\alpha\mu} x^\nu$. 
Notice that $T^{\mu\nu}$ has separate gauge invariant contributions from quarks and gluons \cite{Ji:1996ek},
along with their interaction through the gauge-covariant derivative.
\begin{eqnarray}
 T^{\mu\nu} & = &  T^{\mu\nu}_q + T^{\mu\nu}_g 
             =   \frac{1}{2}[\bar \psi \gamma^{(\mu} i
      \overrightarrow{ D^{\nu)}} \psi + 
       \bar \psi \gamma^{(\mu} i
           \overleftarrow {D^{\nu)} } \psi]  \nonumber \\
           & + &     \frac{1}{4}g^{\mu\nu}
       F^2 - F^{\mu\alpha}F^\nu_{~\alpha}  
\label{e:angmomt}       
\end{eqnarray}


%

The connection of GPDs to the angular momentum becomes apparent by first writing down the matrix element of $T^{\mu\nu}_{q,g}$ for a spin-one system in terms of 
 gravitational form factors as, 
\begin{eqnarray}
& & \langle p' | T^{ \mu \nu} | p \rangle   = 
 -   \frac{1}{2} P^{\mu}P^{\nu} (\epsilon'^*  \epsilon) {\cal G}_{1}(t)
 \nonumber \\ &  -  &
  \frac{1}{4} P^{\mu} P^{\nu} 
\frac{(\epsilon P)(\epsilon'^* P)}{M^2} {\cal G}_{2}(t) 
 -   \frac{1}{2} \left[\Delta^{\mu} \Delta^{\nu} - g^{\mu \nu}
{\Delta^2}\right](\epsilon'^*  \epsilon)\nonumber \\
& \times & {\cal G}_{3}(t)  - \frac{1}{4}
\left[\Delta^{\mu}
\Delta^{\nu} - g^{\mu \nu}{\Delta^2}\right]
\frac{(\epsilon P)(\epsilon'^* P)}{M^2} {\cal G}_{4}(t) 
\nonumber \\
& + & \frac{1}{4} \left[ \left(\epsilon'^{* \mu} (\epsilon P)
 + \epsilon^{\mu} (\epsilon'^* P) \right) P^{\nu} +
\mu \leftrightarrow \nu \right] {\cal G}_{5}(t)
 \nonumber \\
& + & \frac{1}{4} \left[ \left(\epsilon'^{* \mu} (\epsilon P)
 -  \epsilon^{\mu} (\epsilon'^* P) \right) \Delta^{\nu} +
\mu \leftrightarrow \nu  \right. \nonumber \\
& + & \left. 2 g_{\mu\nu} (\epsilon P) (\epsilon'^{'*} P) - (\epsilon'^{* \mu} \epsilon^{\nu}  
+  \epsilon'^{* \nu}\epsilon^{\mu})\Delta^2   \right] {\cal G}_{6}(t) 
\nonumber \\  
& + & 
%
\frac{1}{2} \left[  \epsilon^{* \, \prime \mu} \epsilon^\nu    +   \epsilon'^{* \nu} \epsilon^{\mu}  
\right]  {\cal G}_{7}(t) +  g^{\mu \nu}  (\epsilon^{\prime \, *} \epsilon) M^2 {\cal G}_{8}(t) 
\label{e:angmomdeu}
\end{eqnarray}
where $t = {\Delta}^2$, $P=p+p'$ and $\Delta=p'-p$. There are seven conserved independent form factors, ${\cal G}_{i}(t)$, $i = 1, 7$, and an additional non conserved term,
$g^{\mu \nu}  (\epsilon^{\prime \, *} \epsilon) M^2 {\cal G}_{8}(t)$. In analogy with the nucleon case \cite{Chen:2004cg,Hagler:2004yt}, the enumeration of the independent deuteron EMT form factors, as well its Lorentz structure, was obtained using the partial wave formalism and crossing symmetry (details on our method for counting the form factors are presented in \cite{GGL_odd} (nucleon) and in an upcoming paper \cite{GKLT} (deuteron)).    

Using Eqs.(\ref{e:angmom}) and (\ref{e:angmomdeu}) one can then derive the quark and gluon total angular momentum contribution which reads,
\begin{equation}
J_{q, g}=\frac{1}{2} {\cal G}_{5}(0)   
\end{equation}
One can now connect the gravitational form factors with the coefficients of the correlator for (unpolarized) DVCS. 
For a spin one system one can write this in terms  of five unpolarized GPDs (from the Lorentz symmetric part of the hadronic tensor) \cite{Berger:2001zb},
\begin{eqnarray}
&&  \int \frac{d \kappa}{2 \pi}\,
  e^{i x \kappa P.n}
  \langle p', \lambda' |\,
  \bar{\psi}(-\kappa  n)\, \gamma.n\, \psi(\kappa n)
  \,| p, \lambda \rangle \nonumber \\
& &  =   - (\epsilon'^* .\epsilon) H_1 
+ \frac{(\epsilon .n) (\epsilon'^* .P)+ (\epsilon'^* .n) (\epsilon .P)}{P.n} H_2  \nonumber \\ 
&& - \frac{(\epsilon .P)(\epsilon'^* .P)}{2 M^2} H_3 
+ \frac{(\epsilon .n) (\epsilon'^* .P)- (\epsilon'^* .n) (\epsilon .P)}{P.n} H_4 \nonumber \\ 
&& + \Big\{4 M^2\, \frac{(\epsilon .n)(\epsilon'^* .n)}{(P.n)^2}+\frac{1}{3} (\epsilon'^* .\epsilon) \Big\}H_5 
\label{e:GPDsDeu}
\end{eqnarray}
where $n$ is a light-like vector, and $\epsilon, \epsilon'$ are the polarization vectors of the deuteron in initial 
and final helicity states, respectively.
It follows that  by expanding the matrix element on the left hand side of Eq.(\ref{e:GPDsDeu}) and taking the second moment with respect to $x$ one can find the following relation between the second moments of the GPDs $H_i$ and the form factors ${\cal G}_i$, 
\begin{eqnarray}
&& 2 \! \!\int \!dx x [ H_{1}(x,\xi,t) \!- \frac{1}{3}  H_{5}(x,\xi,t) ] \!=  {\cal G}_{1}(t) + {\xi}^2 {\cal G}_{3}(t)   \nonumber \\
&& \\
&& 2 \! \int dx x H_{2}(x,\xi,t)  =  {\cal G}_{5}(t)   \\ 
&& 2 \! \int dx x H_{3}(x,\xi,t)  =  {\cal G}_{2}(t) + {\xi}^2 {\cal G}_{4}(t)  \\ 
&& - 4 \! \! \int \!dx x H_{4}(x,\xi,t) = \xi {\cal G}_{6}(t) \\
&& \! \! \int dx x H_{5}(x,\xi,t) =   -\frac{t}{8M_D^2} {\cal G}_{6}(t)  + \frac{1}{2} {\cal G}_{7}(t)
\label{e:GPDFFdeu}
\end{eqnarray}
%
In the limit $t \rightarrow 0$ then one finds the sum rule relation between the deuteron GPD $H_2$, and the angular momentum $J_{q, g}$, defined in Eq.(\ref{Ji_deuteron}),
\[
J_{q, g}=\frac{1}{2} \int dx \, x \, H_{2}^{q,g}(x,0,0).
\]
The magnetic moment components of the deuteron current are connected with the $T^{0i}$, components. 
%
%

\begin{figure}
\includegraphics[width=8.cm]{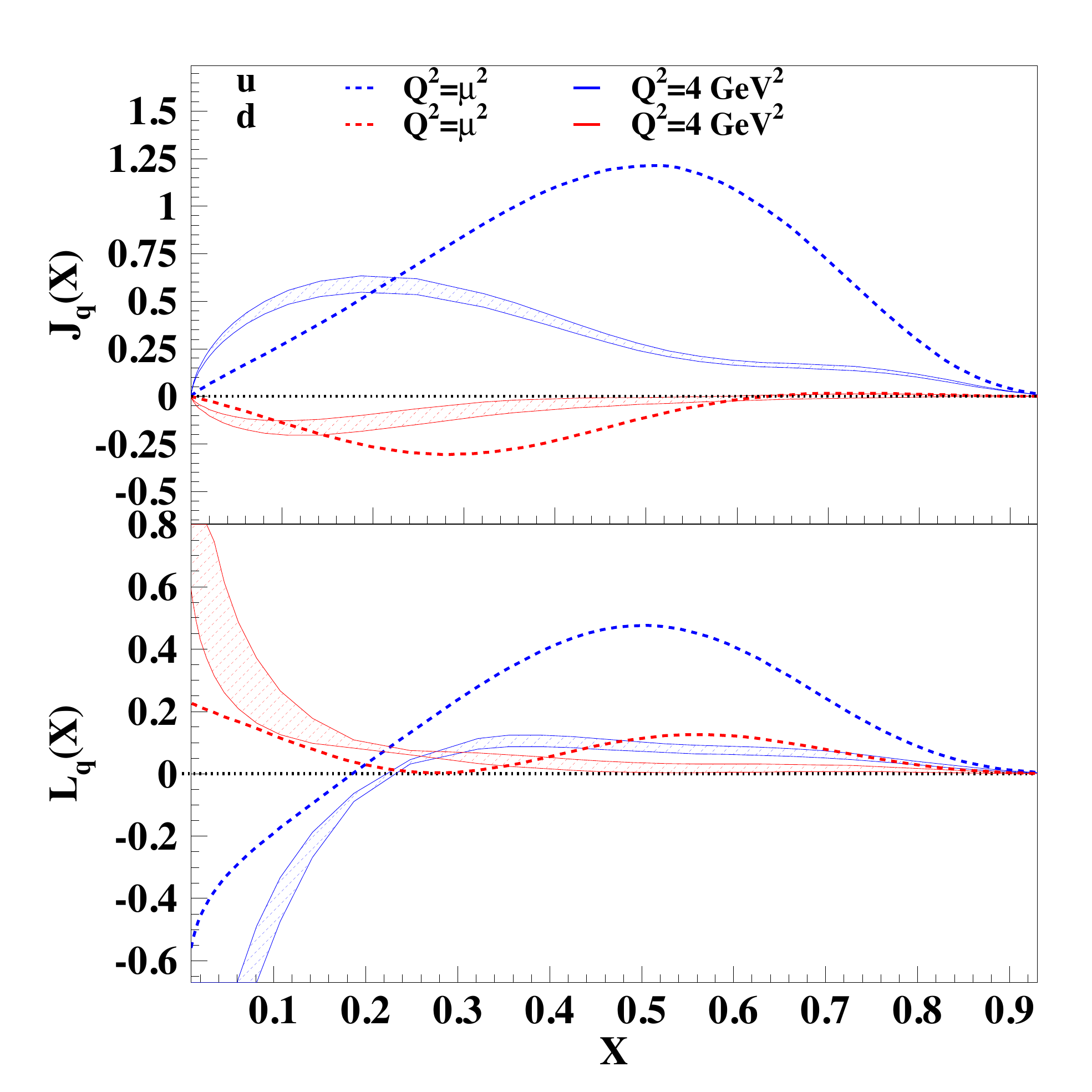}
\caption{(color online) Upper panel: Total angular momentum density distributions, $J_q$, $q=u,d$, calculated using the  GPD parametrization of Ref.\cite{GGL}.  Theoretical error bands are included. Lower panel: Orbital angular momentum density distributions, $L_q$, $q=u,d$, Eq.(\ref{OAM}).  In both panels the dashed lines correspond to the 
scale $\mu^2 \approx 0.1$ GeV$^2$ where spectator models are evaluated \cite{BurkardtBC}; the full lines from our fit results  are calculated at $Q^2= 4$ GeV$^2$. }
\label{fig1}
\end{figure}
The spin one sum rule in Eq.(\ref{Ji_deuteron}), 
which was derived following the same steps as for the spin $1/2$ case, is both the main result and the starting point of our paper.
We now ask the questions: {\it i)} what is the parton content of $H_2$, and ${\it ii)}$ can $H_2$ be extracted from experiment with sufficient 
accuracy? 
In order to explain the partonic sharing of angular momentum in the deuteron 
we start from a picture in terms of bound nucleons.
Eq.(\ref{e:GPDsDeu})  can be written in terms of ``quark-nucleus'' helicity amplitudes 
that depend on $\xi, t$ and $Q^2$ while 
implicitly convoluting over the unobserved quark and nucleon momenta, 
\begin{equation}
C_{\Lambda^\prime \lambda_q^\prime, \Lambda \lambda_q} = \sum_{\lambda_N,  \lambda_N^\prime}  B_{\Lambda^\prime \lambda_N^\prime, \Lambda \lambda_N} \otimes
A_{\lambda_N^\prime  \lambda_q^\prime, \lambda_N  \lambda_q},
\label{h_amps}
\end{equation} 
where $A_{\lambda_N^\prime\lambda_q^\prime;\lambda_N, \lambda_q}$ and $B_{\lambda^\prime,\lambda_N^\prime;\lambda,\lambda_N} $, are the quark-nucleon 
\cite{GGL}, and  nucleon-deuteron helicity amplitudes, respectively, $\Lambda, \lambda_N, \lambda_q$, being the deuteron, nucleon, and quark helicities. 
$H_2$ can be explicitly evaluated from Eq.(\ref{h_amps}) using the convolution formalism that was developed in \cite{LiuTan}, taking care of the angular structure for the deuteron
\cite{CanoPire}.  
For $H_2(x,0,0) = H_2$, 
only the $\{ \Lambda^\prime, \Lambda \} \equiv \{1,1\}$, $\{0,1\}$ deuteron helicity components contribute \cite{CanoPire,GKLT},
\begin{eqnarray}
& H_2& \! \!  =  2 \sum_{\lambda_q}  \left( C_{1 \lambda_q,1 \lambda_q} - \frac{1}{\sqrt{2 \tau_D}} C_{1 \lambda_q,0\lambda_q} \right)
\nonumber \\
& \! \! \approx   &\! \! \! \!  \int\limits_0^{M_D/M} dz  f^{{\bf 1, 1}} (z)  H_N(x/z, 0,0)  + f^{{\bf 0, 1}} (z) 
 E_N(x/z, 0,0), \nonumber \\
\label{parton_struc} 
\end{eqnarray} 
where $H_N=H_u+H_d$, $E_N=E_u+E_d$, are the isoscalar nucleon GPDs, the LC variables,  $x=k^+/(P^+_D/2)$, $z=p^+/(P^+_D/2)$, and $\tau_D=(t_0-t)/2M_D^2$, involve the quark, nucleon and deuteron four-momenta, $k_\mu$, $p_\mu$, and $P_{D, \mu}$, respectively,
\begin{subequations}
\label{fz}
\begin{eqnarray} 
f^{{\bf 1,1}} (z) &  = & \! \! 2 \pi M \! \! \!   \int\limits_{p_{min}(z)}^\infty  dp \; p    \sum_{\lambda}  \chi_{\bf 1}^{* \, \lambda^\prime_{N_1} \lambda_{N_2}}\!  (z,p)  \chi_{\bf 1}^{\lambda_{N_1} \lambda_{N_2}} \!   (z,p)   \nonumber 
\\ && \\
 f^{{\bf 0,1}} (z) &  = &  \! \! 4 \pi M \! \! \int\limits_{p_{min}(z)}^\infty  \! \! \! \!  dp \; p   \sum_{\lambda}  \chi_{\bf 0}^{* \, \lambda^\prime_{N_1} \lambda_{N_2}}\!  (z,p)  \chi_{\bf 1}^{\lambda_{N_1} \lambda_{N_2}} \!   (z,p). \nonumber   \\
\end{eqnarray}
\end{subequations}
where $\lambda_{N_1}$ ($\lambda^\prime_{N_1}$) are the initial (returning) nucleons' helicities, $\lambda_{N_2}$ is the spectator nucleon one, the sum index is $\lambda = \{ \lambda_{N_1}, \lambda^\prime_{N_1}, \lambda_{N_2}\}$; $\chi_{\Lambda}^{\lambda_{N_1}, \lambda_{N_2}}(z,p)$ is the deuteron wave function  \cite{GarVan,GilGro}, 
\begin{eqnarray}
\label{deut_func}
 \chi_{\Lambda}^{\lambda_{N_1},\lambda_{N_2}}(z,p) \! \! & = & \! \! {\cal N} \! \! \sum_{L,m_L,m_S} 
\! \! \left(
 \begin{array}{ccc} 
 j_1 &  j_2 & 1  \\  
 \lambda_{N_1} & \lambda_{N_2} & m_S
 \end{array} \right)   
\! \!  \left(
 \begin{array}{ccc} 
 L &  S & J  \\
 m_L  &  m_S & \Lambda
 \end{array}  \right)  
\nonumber \\ 
& \times & Y_{L \,m_L}\left(\frac{{\bf p}}{p}=\frac{M(1-z)-E}{p} \right) u_L(p),   \nonumber
\end{eqnarray}
where we changed the integration variables from $p_\perp$ to $p = \mid {\bf p} \mid$, therefore a $z$-dependent integration limit appears: $p_{min}(z)=\mid M(1-z) - E \mid$, $M$ being the nucleon mass and $E$ the deuteron's binding energy.      
All formulae are taken in the asymptotic limit, $x$ fixed and $Q^2 \rightarrow \infty$, and the $k_\perp$ dependence is trivially integrated over, in other words no off-shell effects are considered \cite{LiuTan}.  
%
\begin{figure}
\includegraphics[width=8.cm]{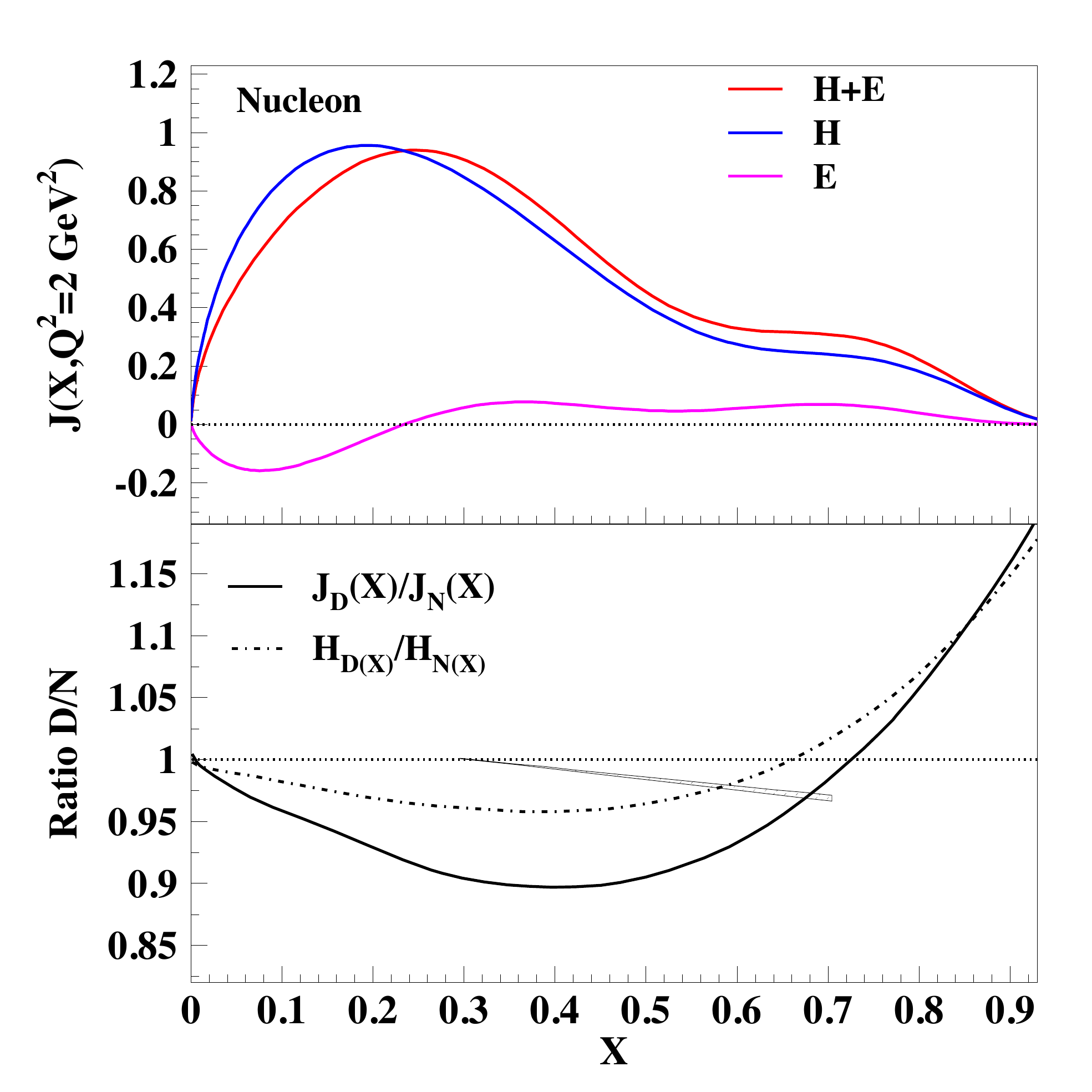}
\caption{(color online) (Upper panel) Contributions $H+E$, $H$, and $E$, to the integrand in the angular momentum sum rule, Eq.(\ref{Ji_deuteron}). All curves were calculated at the scale $Q^2= 4$ GeV$^2$, using the 
parametrization from Ref.\cite{GGL} for the free nucleon. (Lower panel)  The ratio of nuclear to nucleon contributions to angular momentum,  
$H_D/H_N$ (dashes), and  $H_2/(H_N+E_N)\equiv J_D(X)/J_N(X)$, (full curve), calculated using Eq.(\ref{parton_struc}) for the deuteron.  
The small hatched area represents the experimental results from Ref.\cite{Wein}.}
\label{fig2}
\end{figure}

Our results are shown in Figures \ref{fig1} and \ref{fig2}. In Fig.\ref{fig1} we present the proton $u$ and $d$ quarks components of both the total angular momentum density (upper panel), and  the orbital angular momentum density (lower panel),
\begin{equation}
\label{OAM}
L_q(x) = J_q(x) - \frac{1}{2} \Delta q(x), 
\end{equation}
$\Delta q(x)$ being the quark polarized density, and $J_q(X)$ being the integrand in Eq.(\ref{Ji}).  Both the unpolarized and polarized $u$ and $d$ quarks  GPDs used in the calculation are from the parametrization of  Ref.\cite{GGL}. 
The importance of perturbative QCD evolution is evident from the comparison of results at an initial low scale used {\it e.g.} in spectator models, $Q^2 = \mu^2 \approx 0.1 $ GeV$^2$, and 
evolved to $Q^2 = 4$ GeV$^2$ (see discussion in \cite{Thomas}). As a consequence of the Regge behavior of $\Delta q$, the OAM density is peaked at low $x$. 
Our values for the protons angular momentum components are: $J_u = 0.286 \pm 0.011$, $J_d = -0.049 \pm 0.007$, $L_u = -0.104 \pm 0.087$, $L_d = 0.088 \pm 0.031$ at $Q^2=4$ GeV$^2$. 

The total angular momentum density of quarks in  the deuteron is compared to the nucleon one in Fig.\ref{fig2}. The upper panel shows the isoscalar combination,
$J_N(X)=J_u(X)+J_d(X)$ at $Q^2=4$ GeV$^2$.
In the absence of nuclear effects, {\it i.e.} if the deuteron were treated as two independently moving nucleons, in Eq.(\ref{parton_struc}), $f^{11} (z)=f^{01} (z)=\delta(1-z)$, and $H_2=H+E$. Even including nuclear effects, the deuteron angular momentum is dominated by the GPD $H$. 
The separate dependences of the various components in the deuteron, and their impact on OAM are illustrated in the lower panel of Fig.\ref{fig2}, representing 
 the ratio of the nuclear to nucleon contributions to angular momentum,  
$H_D/H_N$ (dashes), and  $H_2/(H_N+E_N)\equiv J_D(X)/J_N(X)$, (full curve). 
As in the forward  case  \cite{HooJafMan}, we find that the D-wave component plays a non trivial role (more details will be given in \cite{GKLT}) producing a most striking angular dependence through the GPD $E$. Its impact is however suppressed. A similar angular dependence can be also shown for $H_5(x,0,0) \equiv b_1$, in agreement with the model calculations of  \cite{HooJafMan}.

How does this affect the spin sum rule? On one side, in a deuteron target, in a deuteron target we observe that
the angular momentum is dominated by the GPD $H$.  If the nuclear effects were found to be small, as predicted within a ``standard" nuclear model, -- nucleons  bound by exchanged mesons -- the deuteron target
would provide an easier access to total angular momentum.  
On the other side, any deviation from the standard nuclear model predictions presented here would signal a different origin of OAM, perhaps related to gluon components, and would therefore be extremely interesting.  
The question of whether the quarks OAM can actually be measured for a deuteron target  is therefore mandatory. While observables were presented in \cite{KirMuel} 
that contain several deuteron GPDs, none of them is sensitive to $H_2$. Here we suggest to measure the deuteron target transverse spin asymmetry, 
$A_{UT}$, which we derive in terms of GPDs as,
\begin{eqnarray}
A_{UT} \approx  -\frac{4\sqrt{\tau_0}}{\Sigma} 
 {\Im}m \! \! \left[ {\cal H}_1^*{\cal H}_5 + \! \! \left( {\cal H}_1^* + \frac{1}{6} {\cal H}_5^*\right)\left({\cal H}_2 - {\cal H}_4 \right) \! \right] \nonumber \\
 \end{eqnarray}
where $\tau_0=\tau(\xi=0)$, $\Sigma$ is the sum of the transversely polarized target cross sections, and ${\cal H}_i$, are the Compton form factors
for the corresponding GPDs. One can see that the term containing ${\cal H}_2$ should dominate the asymmetry, given the expected smallness
of ${\cal H}_5$ \cite{Berger:2001zb,CanoPire}.

\vspace{-0.1cm}
In conclusion, we analyzed the question of OAM in a spin one hadronic system. We derived a sum rule whereby the second moment of the GPD $H_{2}$ 
gives the total angular momentum, $H_{2}$ being the same GPD whose first moment gives the magnetic moment.  
Nuclear effects evaluated within a standard model for the deuteron give $H_2 \approx H+E$, that is OAM in the deuteron 
is predicted to be similar to the sum of the neutron plus proton taken alone. This cancellation is consistent with the smallness of 
the deuteron magnetic moment, reflecting the approximate cancellation between the proton and neutron magnetic moments. 
If found in experiment, deviations from this standard behavior which is calculable to high precision and under control, 
could be a signal of other degrees of freedom such as six quark components, or $k_\perp$ dependent 
re-interactions beyond the collinear convolution considered here.
In either situation studying spin one hadronic systems might shed light on the elusive gluon angular momentum components.  
Finally, we show that measuring angular momentum in the deuteron can be at reach in future experimental facilities
with high enough energy and luminosity,  through transverse spin observables.      

\vspace{-0.2cm}
We thank Lech Szymanowski and Bernard Pire for discussions during the initial stages of this project. 
We are also indebted to Don Crabb for clarifications on deuteron targets polarization, and to Elliott Leader
for many invaluable exchanges.

This work was supported by the U.S. Department
of Energy grants (S.T.) DE-FG02-01ER4120 (K.K., S.L.),  DE-FG02-92ER40702  (G.R.G.).


\end{document}